\documentclass{article}

\usepackage{graphicx}  
\usepackage{amsmath,amssymb} 
\usepackage{algorithm,algorithmic}

\newtheorem{thm}{Theorem}[subsection]
 
 \newtheorem{lem}[thm]{Lemma}
 \newtheorem{prop}[thm]{Proposition}

 \numberwithin{equation}{subsection}

\title{An Alternating l1 approach to the compressed sensing problem}
\author{St\'ephane~Chr\'etien 
\thanks{S. Chr\'etien is with the Laboratoire de Math\'ematiques, UMR CNRS 6623 and Universit\'e de Franche Comt\'e, 
16 route de Gray, 25030 Besan{\c c}on Cedex, France. Email: stephane.chretien@univ-fcomte.fr}}

\begin{document}

\maketitle

\begin{abstract}
Compressed sensing is a new methodology for constructing sensors which allow sparse signals to be efficiently recovered 
using only a small number of observations. The recovery problem can often be stated as the one of finding the solution of an underdetermined system 
of linear equations with the smallest possible support. The most studied relaxation of this hard combinatorial problem is the $l_1$-relaxation consisting 
of searching for solutions with smallest $l_1$-norm. In this short note, based on the ideas of Lagrangian duality, 
we introduce an alternating $l_1$ relaxation for the recovery problem enjoying higher recovery rates in practice than the plain $l_1$ 
relaxation and the recent reweighted $l_1$ method of Cand\`es, Wakin and Boyd. 
\end{abstract}


%

\section{Introduction}
%
%
%
%
Compressed Sensing (CS) is a very recent field of fast growing interest and whose impact on concrete applications in 
coding and image acquisition is already remarkable. Up to date informations on this new topic may be obtained 
from the website {\em http://nuit-blanche.blogspot.com/}. The foundational paper is \cite{Candes:IEEEIT06} where 
the main problem considered was the one of reconstructing a signal from a few frequency measurements. Since then, important 
contributions to the field have appeared; see \cite{Candes:ICM06} for a survey and references therein.

\subsection{The Compressed Sensing problem}
In mathematical terms, the problem can be stated as follows. Let $x$ be a $k$-sparse vector in $\mathbb R^n$, i.e. 
a vector with no more than $k$ nonzero components. The observations are simply given by 

\begin{equation}
\label{linsys}
y=Ax
\end{equation}
where $A\in \mathbb R^{m\times n}$ and $m$ small compared to $n$ with ${\rm rank} A=m$, and the goal is to recover $x$ exactly from these 
observations. The main challenges concern the construction of observation matrices $A$ which allow to 
recover $x$ with $k$ as large as possible for given values of $n$ and $m$. 

The problem of compressed sensing can be solved unambiguously if there is no sparser solution to the linear 
system (\ref{linsys}) than $x$. Then, recovery is obtained by simply finding the sparsest solution to (\ref{linsys}).
If for any $x$ in $\mathbb R^n$ we denote by $\|x\|_0$ the $l_0$-norm of $x$, i.e. the cardinal of the set of indices of 
nonzero components of $x$, the compressed sensing problem is equivalent to 
\begin{equation}
\label{l0}
\min_{x\in \mathbb R^n} \|x\|_0 \hspace{.3cm} {\rm  s.t. } \hspace{.3cm} Ax=y. 
\end{equation}
We denote by $\Delta_0(y)$ the solution of problem (\ref{l0}) and 
$\Delta_0(y)$ is called a decoder \footnote{In the general case where $x$ is not the unique sparsest solution of (\ref{l0}) 
using this approach for recovery is of course possibly not pertinent. Moreover, in such a case, this problem has several solutions with equal 
$l_0$-"norm" and one may rather define $\Delta_0(y)$ as an arbitrary element of the solution set.}. Thus, the CS problem may 
be viewed as a combinatorial optimization problem. Moreover, the following lemma is well known.
\begin{lem}
\label{algcond} 
{\rm (See for instance \cite{Cohen:Pre06})}
If $A$ is any $m \times n$ matrix and $2k \leq  m$, then the following properties are equivalent:

i. The decoder $\Delta_0$ satisfies $\Delta_0(Ax) = x$, for all $x \in \Sigma_k$,

ii. For any set of indices $T$ with $\#T = 2k$, the matrix $A_T$ has rank $2k$ where $A_T$ stands for the submatrix of $A$ composed of the columns 
indexed by $T$ only.
\end{lem}

\subsection{The $l_1$ relaxation}
The main problem in using the decoder $\Delta_0(y)$ for given observations $y$ is that the optimization problem (\ref{l0}) is NP-hard and cannot 
reasonably be expected to be solved in polynomial time. In order to overcome this difficulty, the original decoder $\Delta_0(y)$ has to be replaced 
by simpler ones in terms of computational complexity. Assuming that $A$ is given, two methods have been studied for solving the compressed sensing problem. 
The first one is the orthognal matching pursuit (OMP) and 
the second one is the $l_1$-relaxation. Both methods are not comparable since OMP is a greedy algorithm with 
sublinear complexity and the $l_1$-relaxation offers usually better performances in terms of recovery 
at the price of a computational complexity equivalent to the one of linear programming. More precisely, the $l_1$ 
relaxation is given by 
\begin{equation}
\label{l1}
\min_{x\in \mathbb R^n} \|x\|_1 \hspace{.3cm} {\rm  s.t. } \hspace{.3cm} Ax=y. 
\end{equation}
In the following, we will denote by $\Delta_1(y)$ the solution of the $l_1$-relaxation (\ref{l1}).
From the computational viewpoint, this relaxation is of great interest since it can be solved in 
polynomial time. Indeed, (\ref{l1}) is equivalent to the linear program 
\begin{equation}
\nonumber
\min_{x\in \mathbb R^n} \sum_{i=1}^n z_i \hspace{.3cm} {\rm  s.t. } \hspace{.3cm} -z\leq x\leq z,\hspace{.3cm}{\rm and}\hspace{.3cm} Ax=y. 
\end{equation}
The main subsequent problem induced by this choice of relaxation is to obtain easy to verify sufficient conditions on $A$ for 
the relaxation to be exact, i.e. to produce the sparsest solution to the underdetermined system (\ref{linsys}). A nice condition 
was given by Candes, Romberg and Tao \cite{Candes:IEEEIT06} and is called the Restricted Isometry Property. Up to now, this condition 
could only be proved to hold with great probability in the case where $A$ is a subgaussian random matrix. Several algorithmic approaches have 
also been recently proposed in order to garantee the exactness of the $l_1$ relaxation such as in \cite{Juditsky:ArXiv08} and \cite{Daspremont:ArXiv08}.  
The goal of our paper is different. Its aim is to present a new method for solving the CS problem generalizing the original $l_1$-relaxation of 
(\cite{Candes:IEEEIT06}) and with much better performance in pratice as measured by success rate of recovery versus original sparsity $k$.

\section{Lagrangian duality and relaxations}

\subsection{Equivalent formulations of the recovery problem}
Recall that the problem of exact reconstruction of sparse signals can be solved using $\Delta_0$ and Lemma \ref{algcond}. Let us 
start by writing down problem (\ref{l0}), to which $\Delta_0$ is the solution map, as the following equivalent problem 
\begin{equation}
\label{quad}
\max_{z,\: x\in \mathbb R^n} e^t z 
\end{equation}
subject to 
\begin{equation}
\nonumber
z_ix_i=0, \hspace{.3cm} z_i(z_i-1)=0 \hspace{.3cm} i=1,\ldots,n, \textrm{ and }  Ax=y  
\end{equation}
where $e$ denotes the vector of all ones. Here since the sum of the $z_i$'s is maximized, the variable $z$ plays the role of an indicator function for the event that $x_i=0$. This problem is clearly nonconvex due to the quadratic equality constraints $z_ix_i=0, \hspace{.3cm} i=1,\ldots,n$.

\subsection{The standard Semi-Definite Programming (SDP) relaxation scheme}
\label{SDP}
A simple way to construct a SDP relaxation is to homogenize the quadratic forms in the formulation at hand using a binary variable $z_0=1$. Indeed, 
by symmetry, it will suffice to impose $z_0^1=1$ since, if the relaxation turns out to be exact and a solution $(z_0,z,x)$ is recovered with $z_0=-1$,
then, as the reader will be able to check at the end of this section, $(-z_0,-z,-x)$ will also solve the relaxed problem.  
For instance, problem (\ref{quad}) can be expressed as 
\begin{equation}
\label{quadhom}
\max_{z,\: x\in \mathbb R^n} e^t zz_0 
\end{equation}
subject to 
\begin{equation}
\nonumber
z_ix_i=0, \hspace{.3cm} z_i(z_i-z_0)=0  \textrm{ and }  z_0Ax=y  
\end{equation}
for $i=1,\ldots,n, z_0^2=1$.

If we choose to keep explicit all the constraints in problem (\ref{quadhom}), the Lagrange function can be easily be written as 
\begin{equation}
\nonumber
\begin{array}{l}
L_{SDP}(w,\lambda ,\mu,\nu)=w^tQw+\sum_{i=1}^n \lambda_i w^tC_iw \\
\hspace{1cm}+\sum_{i=1}^n \mu_i w^tE_iw+v_0w^tE_0w\\
\hspace{1cm}+\sum_{j=1}^m \nu_j w^t A_j w-\nu^ty, 
\end{array}
\end{equation}
where $w$ is the concatenation of $z_0$, $z$, $x$ into one vector, $\lambda$ (resp. $\mu$ and $\nu$) is the vector of Lagrange multipliers associated to the constraints 
$z_ix_i=0$, $i=1,\ldots,n$ (resp. $z_i(z_i-z_0)$, $i=1,\ldots,n$, and $z_0a_j^tx=y_j$, $j=1,\ldots,m$) and where all the matrices $Q$, $A_j$, $j=1,\ldots,m$, $E_i$, $i=1,\ldots,n$ and $C_i=1,\ldots,n$ belong to $\mathbb S_{2n+1}$, the set of symmetric $2n+1\times 2n+1$ real matrices and are defined by 
\begin{equation}
\nonumber
Q=
\left[
\begin{array}{ccc}
0 & \frac12 e^t & 0_{1,n} \\
\frac12 e & 0_{n,n} & 0_{n,n} \\
0_{n,1} & 0_{n,n} & 0_{n,n} 
\end{array}
\right]
A_j=
\left[
\begin{array}{ccc}
0 & 0_{1,n} & \frac12 a_j^t \\
0_{n,1} & 0_{n,n} & 0_{n,n} \\
\frac12 a_j & 0_{n,n} & 0_{n,n} 
\end{array}
\right]
\end{equation}
for $j=1,\ldots,m$, where $a_i^t$ is the $j^{th}$ row of $A$, 
\begin{equation}
\nonumber
E_0=
\left[
\begin{array}{ccc}
1 & 0_{1,n} & 0_{1,n} \\
0_{n,1} & 0_{n,n} & 0_{n,n} \\
0_{n,1} & 0_{n,n} & 0_{n,n} 
\end{array}
\right],
E_i=
\left[
\begin{array}{ccc}
0 & -e_i^t & 0_{1,n} \\
-e_i & 2D(e_i) & 0_{n,n} \\
0_{n,1} & 0_{n,n} & 0_{n,n} 
\end{array}
\right]
\end{equation}
and
\begin{equation}
\nonumber
C_i=
\left[
\begin{array}{ccc}
0 & 0_{1,n} & 0_{1,n} \\
0_{n,1} & 0_{n,n} & D(e_i) \\
0_{n,1} & D(e_i) & 0_{n,n} 
\end{array}
\right]
\end{equation}
for $i=1,\ldots,n$ where $e_i$ is the vector with all components equal to zero except the $i^{th}$ which is set to one, $e$ is the vector 
of all ones, $D(e_i)$ is the diagonal matrix with diagonal vector $e_i$ and where $0_{k,l}$ denotes the $k\times l$ matrix of all zeros. The dual function is given by 
\begin{equation}
\nonumber
\theta_{SDP}(\lambda,\mu,\nu)=\sup_{w\in \mathbb R^{2n+1}}L(w,\lambda,\mu,\nu), 
\end{equation}
and thus 
\begin{equation}
\nonumber
\theta_{SDP}(\lambda,\mu,\nu)=\begin{cases}
-\nu^ty \textrm{ if } Q(\lambda,\mu,\nu)\preceq 0 \\
+\infty \textrm{ otherwise }
\end{cases}
\end{equation}
with 
\begin{equation}
\nonumber
Q(\lambda,\mu,\nu)=w^tQw+\sum_{i=1}^n \lambda_i w^tC_iw+\sum_{i=0}^n \mu_i w^tE_iw+\sum_{j=1}^m \nu_j w^t A_j w
\end{equation}
and where $\succeq$ is the L\"owner ordering ($A\succeq B$ iff $A-B$ is positive semi-definite). 
Therefore, the dual problem is given by 
\begin{equation}
\nonumber
\inf_{\lambda \in \mathbb R^n, \mu\in \mathbb R^{n+1}, \nu\in \mathbb R^m} \theta_{SDP}(\lambda,\mu,\nu),
\end{equation}
which is in fact equivalent to the following semi-definite program 
\begin{equation}
\label{dualsdp}
\inf_{\lambda \in \mathbb R^n, \mu\in \mathbb R^{n+1}, \nu\in \mathbb R^m} -y^t \nu,
\end{equation}
subject to 
\begin{equation}
Q(\lambda,\mu,\nu)\preceq 0. 
\end{equation}
We can also try and formulate the dual of this semi-definite program which is called the bidual of the initial problem. This bidual problem is easily seen after some 
computations to be given by 
\begin{equation}
\label{bidual}
\max_{X\in \mathbb S_{2n+1},\: X\succeq 0} {\rm trace} (QX) 
\end{equation}
subject to 
\begin{equation}
\nonumber
{\rm trace} (A_jX)=y_j,\: j=1,\ldots,m,  
\end{equation}
\begin{equation}
\label{z0}
{\rm trace}(E_0X)=1, 
\end{equation}
\begin{equation}
\nonumber
{\rm trace}(E_iX)=0 \textrm{ and } {\rm trace}(C_iX)=0,\: i=1,\ldots,n. 
\end{equation}
Now, if $X^*$ is an optimal solution with ${\rm rank}(X^*)=1$, then 
\begin{equation}
\nonumber
X^*=\Big(\pm
\left[
\begin{array}{c}
z_0^* \\
z^* \\
x^*
\end{array}
\right]\Big)
\Big(\pm 
\left[
\begin{array}{c}
z_0^* \\
z^* \\
x^*
\end{array}
\right]\Big)^t 
\end{equation}
and it can be easily verified that all the constraints in (\ref{quadhom}) are satisfied. Moreover, we may additionally impose that $z_0^*=1$ \footnote{Indeed, if $z_0^*=-1$, 
multiply by $-1$ the whole vector $[z_0^*,z^*,x^*]$}. However, the following proposition ruins the hopes for the occurance of such an agreable situation.  
\begin{prop}
\label{optrank}
If non empty, the solution set of the bidual problem (\ref{bidual}) is not a singleton and it contains matrices with rank equal to $n-m$.  
\end{prop}
{\bf Proof}. 
Consider the subspace $W_0$ of $\mathbb R^{2n+1}$ as the set of vectors whose $n+1$ first coordinates are equal to zero and such that the last $n$ coordinates 
form a vector in the kernel of $A$. Since we assumed that ${\rm rank} A=m$, we have that ${\rm dim} W_0=n-m$. 
Assume that  there exists a solution $X^*$ to (\ref{bidual}) with rank less than or equal to $n-m-1$. Then, it is possible to find a vector $w$ in $W_0$ with $w^t\perp P_{W_0}({\rm Range}(X^*))$.
On the other hand, one can easily check 
that $X^{**}=X^*+ww^t$ satisfies all the bidual constraints and has the same objective value as $X^*$. Thus, 
$X^{**}$ is also a solution of the bidual problem  and ${\rm rank} X^{**}={\rm rank} X^*+1$. Iterating the argument up to matrices of dimension equal to $n-1$, we obtain that the solution set 
contains matrices with rank equal to $n-m$. To prove non uniqueness of the solution, 
for any solution matrix $X^*$, set $X^{***}=X^*+ww^t$ for any choice of $w$ in $W_0$ and $X^{***}$ is also a solution of the bidual problem. 
\hfill$\Box$ 

\subsection{Comments on the SDP relaxation}
Despite the powerfull Lagrangian methodology behind its construction, the SDP relaxation of the problem has three major drawbacks:
\begin{itemize}
\item as implied by Proposition \ref{optrank}, the standard SDP relaxation scheme leads to solutions which naturally have rank greater than one which makes it hard to try and 
recover a nice primal candidate. Moreover, even if the rank problem could be overcome in practice in the case where $x$ is sparse enough, by adding more ad hoc constraints in the SDP, finding the most natural way to do this seemed quite non trivial to us.  
\item in the case where the SDP has a duality gap, proposing a primal suboptimal solution does not seem to be an easy task. 
\item the computational cost of solving Semi-Definite Programs is much greater than the cost of solving our naive relaxation, a fact which may be important in real applications. 
\end{itemize}

\subsection{An utopic relaxation}
\label{Utop}
In order to overcome the drawbacks of the SDP relaxation, we investigate another scheme which may look utopic at first sight. 
Notice that one interesting variant of formulation (\ref{quad}) could be the following in which the
nonconvex complementarity constraints are merged into the unique constraint $\|D(z)x\|_1=0$ 
\begin{equation}
\label{prel1}
\max_{z\in \{0,1\}^n} e^t z \hspace{.3cm} {\rm  s.t. } \|D(z)x\|_1=0, \hspace{.3cm} Ax=y.
\end{equation}
Choosing to keep the constraints $Ax=y$ and $z\in \{0,1\}^n$ implicit in (\ref{prel1}), the Lagrangian function 
is given by 
\begin{equation}
L(x,z,u)=e^tz-u\|D(z)x\|_1
\end{equation}
where $D(z)$ is the diagonal matrix with diagonal vector equal to $z$.
The dual function (with values in $\mathbb R\cup +\infty$) is defined by 
\begin{equation}
\label{theta}
\theta(u)=\max_{z\in \{0,1\}^n, \: Ax=y} L(x,z,u)
\end{equation}
and the dual problem is 
\begin{equation}
\label{dual}
\inf_{u\in \mathbb R} \theta(u).  
\end{equation}
The main problem with the dual problem (\ref{dual}) is that the solutions to (\ref{theta}) are as difficult to 
obtain as the solution of the original problem (\ref{prel1}) because of the nonconvexity of the Lagrangian function $L$.

\section{The Alternating $l_1$ method}
We now present a generalization of the $l_1$ relaxation which we call the Alternating $l_1$ relaxation with better experimental 
performances than the standard $l_1$ relaxation and the SDP relaxation. 

\subsection{A practical alternative to the utopic relaxation}
 
Due to the difficulty of computing the dual function $\theta$ in the relaxation \ref{Utop}, the interest of this scheme seems at first to be of pure theoretical nature only. 
In this section, we propose a suboptimal but simple alternating minimization approach.

When we restrict $z$ to the value $z=e$, i.e. the vector of all ones, solving the problem
\begin{equation}
\label{partialtheta}
x_(u)={\rm argmax}_{z=e, \: x\in \mathbb R^n, \: Ax=y} L(x,z,u)
\end{equation}
gives exactly the solution $\Delta_1(y)$ of the $l_1$ relaxation. From this remark, and the Lagrangian duality theory 
above, it may be supected that a better relaxation can be obtained by trying to optimize the Lagrangian even in a suboptimal manner. 

\vspace{.3cm}

\begin{algorithm}
\caption{Alternating $l_1$ algorithm (Alt-$l_1$)}
\begin{algorithmic}
\REQUIRE $u>0$ and $L \in \mathbb N_*$         

\STATE $z_u^{(0)}=e$
\STATE $x_u^{(0)}\in{\rm argmax}_{x\in \mathbb R^n, \: Ax=y} L(x,z^{(0)},u)$
\STATE $l=1$
\WHILE {$l\leq L$}
   \STATE $z_u^{(l)}\in {\rm argmax}_{z\in \{0,1\}^n} L(x_u^{(l)},z,u)$
   \STATE $x_u^{(l)}\in {\rm argmax}_{x\in \mathbb R^n, \: Ax=y} L(x,z_u^{(l)},u)$       
   \STATE $l \leftarrow l+1$
\ENDWHILE

\STATE Output $z_u^{(L)}$ and $x_u^{(L)}$.
\end{algorithmic}
\end{algorithm}

At each step, knowing the value of $z_u^{(l)}$ implies that optimization with respect to $x\in\mathbb R^n$
can be equivalently restricted to the set of variables $x_i$ which are indexed by the $i$'s associated with the values of $z_u^{(l)}$ which are equal to one.
Thus, the choice of $z_u^{(l)}$ corresponds to adaptive support selection for the signal to recover.  

The following lemma states that $z_u^{(l)}$ is in fact the solution of a simple thresholding procedure.
\begin{lem}
\label{01}
For all $x$ in $\mathbb R^n$, any solution $z$ of 
\begin{equation}
\label{rlxstp1}
\max_{z\in [0,1]^n} L(x,z,u) 
\end{equation}
satisfies that $z_i=1$ if $|x_i|< \frac1{u}$, 0 if $|x_i|> \frac1{u}$ and
$z_i\in [0,1]$ otherwise. 
\end{lem}

{\bf Proof}.
Problem (\ref{rlxstp1}) is clearly separable and the solution can be easily computed coordinatewise. 
\hfill$\Box$
\section{Monte Carlo experiments}
In this section, using Monte Carlo experiments, we compare our Alternating $l_1$ approach 
to two recent methods proposed in the litterature: the Reweighted 
$l_1$ of Cand\`es, Wakin and Boyd \cite{Candes:JFAA08}  and the Iteratively Reweighted Least-Squares as proposed in \cite{Chartrand:ICASSP08}. 
The problem size was chosen to be the same as in Chartrand and Yin's paper \cite{Chartrand:ICASSP08}: $n=256$, $m=100$. 
For each sparsity $k$ level a hundred different $k$-sparse vectors $x$ were drawn as follows:
the support was chosen uniformly on all support with cardinal $k$ and the nonzero components were drawn from the Gaussian distribution
$\mathcal N(0,4)$. The,n, the observation matrix was obtained in two steps: first draw a $m\times n$ matrix with i.i.d. Gaussian 
$\mathcal N(0,1)$ entries and then normalize each column to 2 as in \cite{Chartrand:ICASSP08}. 

The parameter $u$, namely the Lagrange multiplier for the complementarity constraint was tuned as follows: since on the one hand the natural breakdown point 
for $l_0/l_1$ equivalence, i.e. equivalence of using $l_0$ vs. $l_1$ minimization, lies around $k=\frac{m}4$ and on the other hand, the Alternating 
$l_1$ is nothing but a successive thresholding algorithm due to Lemma \ref{01}, we decided to chose the smallest possible $u$ so that the 
$\frac{m}4$ largest components $x_u^{(0)}$ the first step of the Alternating $l_1$ algorithm (which is nothing but the plain $l_1$ decoder whatever the value of $u$) 
be over $\frac1{u}$. Notice that this value of $u$ is surely not the solution of the dual problem but our choice is at least motivated by 
reasonable deduction based on pratical observations whereas the tuning parameter in the other two algorithms is not known to enjoy such 
an intuitive and meaningful selection rule. We chose $L=4$ in these experiments. The numerical results for the IRLS and the Reweighted $l_1$ 
were communicated to us by Rick Chartrand whom we greatly thank for his collaboration. 

Comparison between the success rates the three methods is shown in Figure 1. Our Alternating $l_1$ method outperformed both the Iteratively 
Reweighted Least Squares and the Reweighted $l_1$ methods for the given data size. As noted in \cite{Chartrand:ICASSP08}, the IRLS and the Reweighted $l_1$ 
enjoy nearly the same exact recovery success rates. 

\begin{figure}[htb]
\label{comp}
\begin{center}
\includegraphics[width=8.5cm]{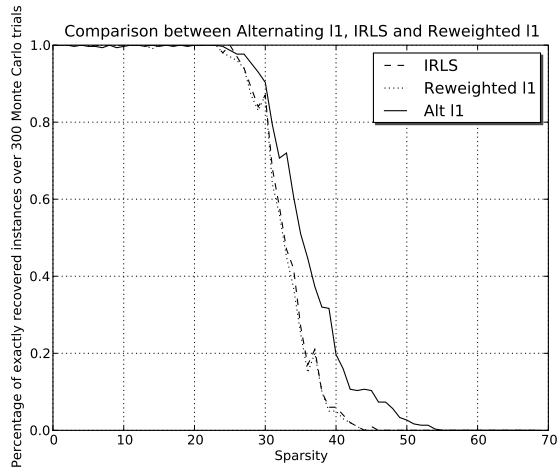}
\caption{Rate of success over 100 Monte Carlo experiments in recovering the support of the signal vs. signal sparsity $k$ for $n=256$, $m=100$, $L=4$. }
\end{center}
\end{figure}

{\bf Remark}. The Reweighted $l_1$ and the 
Reweighted LS both need a value of $\epsilon$ (or even a sequence of 
such values as in \cite{Chartrand:InvProb08}) which is hard to optimize ahead of time, whereas the value $u$ in the Alternating $l_1$ is a Lagrange multiplier, 
i.e. a dual variable. In the Monte Carlo experiments of the previous 
section, we decided to base our choice of $u$ on a simple an intuitive criterion suggested by the well known experimental behavior of the plain $l_1$ relaxation. 
On the other hand, it should be interesting to explore duality a bit further 
and perform experiments in the case where $u$ is approximately optimized (using any derivative free procedure for instance) based on 
our heuristic alternating $l_1$ approximation of the dual function $\theta$.

\end{document}